\documentclass[12pt,a4paper,titlepage]{article}

\usepackage{indentfirst} 
\usepackage{lastpage} 

\usepackage{amsmath} 
\usepackage{amssymb}
\usepackage{amsfonts}
\usepackage{amsthm}
\usepackage{bm} 
\usepackage{icomma} 

\usepackage{extsizes} 
\usepackage{geometry} 
\usepackage{setspace}
\usepackage{textcase}

\usepackage{cancel}

\usepackage[linkcolor=blue,colorlinks=true]{hyperref}

\newtheorem{proposition}{Proposition}
\newtheorem{lemma}{Lemma}
\newtheorem{theorem}{Theorem}
\newtheorem{definition}{Definition}
\newtheorem{corollary}{Corollary}

\theoremstyle{definition}
\newtheorem*{demo}{Proof}

\usepackage{epsfig}
\usepackage{graphicx}

\begin{document}
\begin{center}	
	
	\Large
	\textbf{Symplectic analogs of polar decomposition and their applications to bosonic Gaussian channels}
		
	\large 
	\textbf{A.E. Teretenkov}\footnote{Department of Mathematical Methods for Quantum Technologies,
		Steklov Mathematical Institute of Russian Academy of Sciences, Moscow, Russia\\ E-mail:taemsu@mail.ru}
	
\end{center}

\footnotesize
We obtain several analogs of real polar decomposition for even dimensional matrices. In particular, we decompose a non-degenerate matrix as a product of a Hamiltonian and an anti-symplectic matrix and under additional requirements we  decompose a matrix as a skew-Hamiltonian and a symplectic matrix. We apply our results to study bosonic Gaussian channels up to inhomogeneous symplectic transforms. \\


\normalsize

\section{Introduction}

It is well known that an arbitrary real matrix could be represented as a product of a real non-negative definite symmetric matrix and a real orthogonal matrix, i.e. in ordinary polar decomposition \cite[Sec. IX, \S 14]{Gantmacher77}, \cite[Sec. 7.3]{Horn13}. It was generalized in several ways both for real and complex, both for general and structured matrices with different conditions on factors \cite{Bolshakov99}, \cite[Secs. 5--6]{Mackey05}, \cite{VanDerMee01}, \cite{Higham10}. In this study we consider the analogs of real polar decomposition, in which symplectic or anti-symplectic matrices play the role analogous to the orthogonal one. 

In Sec. \ref{sec:analogs} we present our main results. In theorem \ref{th:main0} we prove that an arbitrary real even-dimensional non-degenerate matrix could be represented as a product of a Hamiltonian and an anti-symplectic matrix. This result is mainly based on \cite{Fass99}. In theorem \ref{th:main} we prove that under certain conditions a real even-dimensional non-degenerate matrix could be represented as a product of a skew-Hamiltonian and a symplectic matrix. This result is mainly based on computation of a primary square root \cite[Ch. 6]{Higham08}, \cite{Higham87}. Several other symplectic analogs for real polar decomposition follow directly from these theorems and are formulated as corollaries \ref{cor:decomRDS}--\ref{cor:last}.

Our research was inspired by the study of the forms to which bosonic Gaussian channels \cite[Sec.~12.4]{Holevo10} could be reduced by inhomogeneous symplectic transforms. In Sec. \ref{sec:bosChan} we present this application of our study in the form which needs the knowledge of linear algebra only rather than quantum information theory. The bosonic Gaussian channels play an important role in various quantum information topics \cite{Holevo01}--\cite{Hol17}. They also naturally arise as exactly solvable models of quantum Markovain dynamics \cite{Manko79}--\cite{Teretenkov19}. The main result could be found in theorem \ref{th:applToBGQC}.

In Conclusions we summarize the results of this work and discuss the possible directions of the future study.

\section{Symplectic analogs of polar decomposition}\label{sec:analogs}

In our work we denote the set of $ 2n \times 2n $ real matrices by  $ \mathbb{R}^{2 n \times 2n} $ and take the matrix of the symplectic form as (we follow the agreement for signs and ordering from \cite[Eq. (1)]{Wolf2008}, \cite[App.~6]{Arnold13})
\begin{equation*}
J = 
\begin{pmatrix}
0 & -I_n\\
I_n & 0
\end{pmatrix},
\end{equation*}
where $ I_n $ is the $ n \times n $ identity matrix. Let us note that $ J = - J^{-1} = - J^{T}$ and $ \det J = 1 $. We also need the following lemma.

\begin{lemma}\label{lem:propOfY}
	Let $ X \in \mathbb{R}^{2 n \times 2n} $, then the matrix
	\begin{equation}\label{eq:matY}
	Y =  - X J X^T J
	\end{equation}
	is skew-Hamiltonian, i.e.  $ (J Y)^T = - J Y$. If, in addition, $ X $ is non-degenerate, then $ Y $ also is. 
\end{lemma}

\begin{demo}
	Let us explicitly calculate  $ Y^T= - J X J X^T =  J (- X J X^T J) J^{-1} = J Y J^{-1} $ and, hence, $ (J Y)^T = - Y^T J =  -  J Y J^{-1} J =  -  J Y$. 
	
	If  $ X $ is non-degenerate, then $ \det X \neq 0 $. So $ \det Y =  \det(- X J X^T J) = (-1)^{2n} (\det X)^2 (\det J)^2 = (\det X)^2 \neq 0$. Thus, $ Y $ is non-degenerate. \qed
\end{demo}

\begin{theorem}\label{th:main0}
	Let $ X \in \mathbb{R}^{2n \times 2n} $ and $ \det X \neq 0 $, then there exists a Hamiltonian matrix $ H $  ($ (J H)^T = J H $) and a real  anti-symplectic matrix $ T $ ($ TJT^T= -J$) such that
	\begin{equation}\label{eq:decomHT}
	X = H T.
	\end{equation}
\end{theorem}

\begin{demo}
	Let us define the matrix $ Y $ by formula \eqref{eq:matY}, then	by lemma \ref{lem:propOfY} it is skew-Hamiltonian and non-degenerate. From \cite{Fass99} their exists a  Hamiltonian square root $ H $, i.e. $ (J H)^T = J H $ and $ H^2 = Y $. $ H $ is also non-degenerate; otherwise, $ \det Y = (\det H)^2 = 0$ and the matrix $ Y $ would be degenerate. Thus, one could define the matrix $ T $ by the formula $ T = H^{-1} X $. Let us prove that $ T $ is anti-symplectic by the direct computation
	\begin{align*}
	T J T^T &=  H^{-1} X J X^T (H^T)^{-1} = H^{-1} X J X^T (J H J)^{-1} = H^{-1} X J X^T (J H J)^{-1} \\
	&= H^{-1} X J X^T J H^{-1} J = H^{-1} (-Y) H^{-1} J = -H^{-1} H^2 H^{-1} J = -J,
	\end{align*}
	where $ H^T = J H J $ is taken into account, which is equivalent to the fact that $ H $ is skew-Hamiltonian. \qed
\end{demo}

Let us note that decomposition \eqref{eq:decomHT} is not unique as the Hamiltonian square root $ H $ of the skew-Hamiltonian matrix $ Y $ is not unique \cite[Sec. 5]{Fass99}.

\begin{corollary}\label{cor:decomRDS}
	Let $ X \in \mathbb{R}^{2n \times 2n} $ and $ \det X \neq 0 $, then there exists a real symmetric matrix $ R $  ($ R^T = R $) and a real  symplectic matrix $ S $ ($ SJS^T= J$) such that
	\begin{equation*}
	X = R D S,
	\end{equation*}
	where
	\begin{equation}\label{eq:matD}
	D = \begin{pmatrix}
	I_n & 0\\
	0	& -I_n
	\end{pmatrix}.
	\end{equation}
\end{corollary}

\begin{demo}
	Let us decompose the matrix $ H $ from theorem \ref{th:main0} as $ H= R J $, then $ R $ is a symmetric matrix. Let us define the matrix
	\begin{equation*}
	Z = \begin{pmatrix}
	0 & -I_n\\
	-I_n &  0
	\end{pmatrix},
	\end{equation*}
	then $ Z J Z^T = - J$ and, hence, $ Z $ is anti-symplectic. So the matrix $ S = Z T $ is a symplectic matrix as a product of two anti-symplectic ones. Let us also note that $ J Z = D $ and $ Z^2 = I_{2n} $. Thus, we have the matrices $ R $, $ D $, $ S $ such that $ R D S=R J Z^2 T= HT =X $ and they have the desired properties.\qed
\end{demo}

\begin{corollary}\label{cor:transposed}
	Let $ X \in \mathbb{R}^{2n \times 2n} $ and $ \det X \neq 0 $, then there exists
	
	(1) a Hamiltonian matrix $ H $  ($ (J H)^T = J H $) and a real  anti-symplectic matrix $ T $ ($ TJT^T= -J$) such that
	\begin{equation*}
	X = T H ;
	\end{equation*}
	
	(2) a real symmetric matrix $ R $  ($ R^T = R $) and a real  symplectic matrix $ S $ ($ SJS^T= J$) such that
	\begin{equation}\label{eq:decompSDR}
	X = S D R,
	\end{equation}
	where the matrix $ D $ is defined by formula \eqref{eq:matD}.
\end{corollary}

The proof consists in application of theorem \ref{th:main0} and corollary \ref{cor:decomRDS} to the matrix $ X^T $ and taking into account that the transpose of Hamiltonian, anti-symplectic, symmetric and symplectic matrices are themselves Hamiltonian, anti-symplectic, symmetric and symplectic, respectively, and $ D^T =D $.

For the formulation of the following theorem the definitions of the primary function and the principal primary square root of a matrix are needed, they could be found in Sec.~1.6 and Sec.~1.7 in \cite{Higham08}, respectively.

\begin{theorem}\label{th:main}
	Let $ X \in \mathbb{R}^{2n \times 2n} $ such that the matrix $ Y =  - X J X^T J $ has no zero or negative real eigenvalues, then there exists a real skew-Hamiltonian matrix $ M $ ($ (J M)^T = -J M $) and a real  symplectic matrix $ S $ ($ SJS^T= J$) such that
	\begin{equation*}
	X = M S,
	\end{equation*}
	where any primary square root of $ Y $ could be taken as $ M $, in particular the principal  primary square root of $ Y $.
\end{theorem}

\begin{demo}
	By theorem 7 in \cite{Higham87} the matrix $ Y $ has a real primary square root $ M $ ($ M^2 =Y $) which is a polynomial of $ Y $, i.e. $ M  = \sum_{k=1}^{2n} a_k Y^k, a_k \in \mathbb{R} $. Hence, if $Y^T=  J Y J^{-1} $, then $ M^T=  J M J^{-1} $.
	
	As $ M $ has no zero eigenvalues, then let us define $ S =M^{-1} X $. By direct computation we verify that $ S J S^T = M^{-1} X J X^T (M^T)^{-1}  =  M^{-1} X J X^T J M^{-1} J^{-1} =  M^{-1} (- M^2) M^{-1} J^{-1} = J $. Thus, the matrix $ S $ is symplectic. 
	
	By theorem 1.29 from \cite[p. 41]{Higham08} the matrix $ M $ could be computed as the principal square root of $ Y $.\qed
\end{demo}

It could also be regarded as a special case of general theorem 6.2 from \cite{Mackey05}. A similar theorem could be found in \cite[Theorem 7]{Ikramov01} but for the matrices over the field $ \mathbb{C} $; in such a case only absence of zero eigenvalues is needed as in our theorem \ref{th:main0}.

\begin{corollary}\label{cor:decompAS}
	Let $ X \in \mathbb{R}^{2 n \times 2n} $ such that the matrix $ Y =  - X J X^T J $ has no zero or negative real eigenvalues, then there exists a real skew-symmetric matrix $ A $ ($ A^T = - A $) and a real  symplectic matrix $ S $ ($ S JS^T= J$) such that
	\begin{equation*}
	X = A S.
	\end{equation*}
\end{corollary}
The proof is based on the fact that $ J $  by itself is a symplectic matrix and assuming  $ A = -M J $, where $ M $ is defined by theorem \ref{th:main}. 

\begin{corollary}\label{cor:decompMDSandADS}
	Let $ X \in \mathbb{R}^{2 n \times 2n} $ such that the matrix $ Y =  - X J X^T J $ has no zero or positive real eigenvalues, then there exists
	
	(1) a real skew-Hamiltonian matrix $ M $ ($ (J M)^T = -J M $) and a real  symplectic matrix $ S $ ($ SJS^T= J$) such that
	\begin{equation*}
	X = M D S;
	\end{equation*}
	
	(2) a real skew-symmetric matrix $ A $ ($ A^T = - A $) and a real  symplectic matrix $ S $ ($ S JS^T= J$) such that
	\begin{equation*}
	X = A D S,
	\end{equation*}
	where $ D $ is defined by formula \eqref{eq:matD}.
\end{corollary}

\begin{demo}
	Let us prove paragraph (2), the proof of paragraph (1) is similar. If $ Y =  - X J X^T J $ has no zero or positive real eigenvalues, then the matrix \\
	\begin{equation*}
	- (X D) J (X D)^T J = - X D J D X^T J  = - Y
	\end{equation*}
	has no zero or negative eigenvalues. Hence, the matrix $ X D $ satisfies the conditions of corollary \ref{cor:decompAS}. So we could represent $  X D = A S_1 $, where $ A= -A^T $ and $ S_1 J S_1^T= J $. Then $  X  = A D (D S_1 D) = A D S $, where $ S =D S_1 D$ is symplectic $ S J S^T = D S_1 D J D S_1^T D  = -D S_1 J S_1^T D = -D J D = J$. \qed
\end{demo}

\begin{corollary}\label{cor:last}
	Let $ X \in \mathbb{R}^{2 n \times 2n} $ such that the matrix $ Y =  - X^T J X J $ has no zero or negative real eigenvalues, then there exists
	
	(1) a real skew-Hamiltonian matrix $ M $ ($ (J M)^T = -J M $) and a real  symplectic matrix $ S $ ($ SJS^T= J$) such that
	\begin{equation*}
	X = S M;
	\end{equation*}
	
	(2)  a real skew-symmetric matrix $ A $ ($ A^T = - A $) and a real  symplectic matrix $ S $ ($ S JS^T= J$) such that
	\begin{equation}\label{eq:decompSA}
	X = S A.
	\end{equation}
	
	 Let $ X \in \mathbb{R}^{2 n \times 2n} $ such that the matrix $ Y =  - X^T J X J $ has no zero or positive real eigenvalues, then
 	there exists
 		
	(3) a real skew-Hamiltonian matrix $ M $ ($ (J M)^T = -J M $) and a real  symplectic matrix $ S $ ($ SJS^T= J$) such that
	\begin{equation*}
	X = S D M;
	\end{equation*}
	
	(4) a real skew-symmetric matrix $ A $ ($ A^T = - A $) and a real  symplectic matrix $ S $ ($ S JS^T= J$) such that
	\begin{equation}\label{eq:decompSDA}
	X = S D A,
	\end{equation}
	where $ D $ is defined by formula \eqref{eq:matD}.
\end{corollary}

The proof is analogous to the one of corollary \ref{cor:transposed} and is a  direct result of application of theorem~\ref{th:main} and corollaries \ref{cor:decompAS}, \ref{cor:decompMDSandADS} to the matrix $ X^T $.

\section{Bosonic Gaussian quantum channels}\label{sec:bosChan}

In spite of the fact that a bosonic Gaussian channel is a certain map in infinite-dimensional Banach space it is fully characterized by the triple consisting of two real matrices and a vector in $ 2n $ dimensional real Euclidean space \cite[Subsec. 12.4.1]{Holevo10}. Thus, we give a pure linear algebraic definition for a Gaussian bosonic channel:
\begin{definition}
	The triple $ (K, l, \alpha) $, where $ K, \alpha \in \mathbb{R}^{2n \times 2n } $, $ l \in   \mathbb{R}^{2n}$, $ \alpha = \alpha^T $, with an additional condition that the  following complex matrix is non-negative definite 
	\begin{equation}\label{eq:GaussCond}
	\alpha - \frac{i}{2} (J - K^T J K) \geqslant 0,
	\end{equation} 
	is called a bosonic Gaussian channel.
\end{definition}
Such a definition is enough for the purposes of our work, because such a triple uniquely defines a bosonic Gaussian channel, which arises in quantum information theory as a map in operator spaces \cite[Theorem 12.30]{Holevo10}.  

Let us formulate the results of \cite[p. 301]{Holevo10} as the following proposition.
\begin{proposition}
	The  bosonic Gaussian channels form a monoid with respect to the product defined by the following formula 
	\begin{equation}\label{eq:prod}
	(K_2, l_2, \alpha_2) (K_1, l_1, \alpha_1) = (K_1 K_2, K_2^T l_1 + l_2,  K_2^T \alpha_1 K_2 + \alpha_2) .
	\end{equation}
\end{proposition}
This means that  if the triples at the left-hand side of  \eqref{eq:prod} are bosonic Gaussian channels, then the right-hand side of it also is, and for $ (K_1, l_1, \alpha_1) = (I_{2n}, 0, 0) $ the right-hand side of  \eqref{eq:prod} equals $ (K_2, l_2, \alpha_2)  $. 

The triples of the form $ (S, l, 0) $, where $ SJS^T =J $, satisfy  \eqref{eq:GaussCond}. They form the group of in\-homo\-geneous (affine) symplectic transformations \cite[Sec. 3.5]{DeGosson16} which, hence, is a submonoid of the monoid of bosonic Gaussian channels. The problem of simplifying a Gaussian channel by multiplying it from the left and from the right by inhomogeneous symplectic transformations naturally occurs in quantum information theory \cite{Holevo07}, \cite{Wolf2008},  \cite{Caruso08}, \cite[Subsec.~12.6.1]{Holevo10}.

Here two strategies could be applied: one could simplify the matrix $ K $ as much as possible without simplifications for the matrix $ \alpha $ or one could simplify the form of the matrix $ \alpha $ and only then try to simplify the matrix $ K $ without change of the simplified form for the matrix $ \alpha $.  The first strategy leads to the Wolf canonical form \cite{Wolf2008}, the latter one could be presented here. We are inspired by the fact that in Holevo canonical forms for one-mode channels \cite{Holevo07}, \cite[Subsec.~12.6.1]{Holevo10}, i.e. for the case when $ n=1 $, the matrix $ \alpha $ is fully diagonalized.

\begin{theorem}\label{th:applToBGQC}
	Let  $ (K, l, \alpha)  $ be a bosonic Gaussian channel such that $ \det K \neq 0 $, then
	
	(1) there exist two in\-homo\-geneous  symplectic transformations $ (S_1, h_1, 0) $ and $ (S_2, h_2, 0) $ such that
	\begin{equation}\label{eq:channelDRform}
	(S_2, h_2, 0)  (K, l, \alpha) (S_1, h_1, 0) = (D R , 0, \Lambda),
	\end{equation}
	where $ R $ is  a real symmetric matrix ($ R = R^T $), $ D $ is defined by formula \eqref{eq:matD} and $ \Lambda $ is  a real diagonal matrix;
	
	(2) if the matrix $ -K^T J K J $ has no negative real eigenvalues, there exist two in\-homo\-geneous  symplectic transformations $ (S_1, h_1, 0) $ and $ (S_2, h_2, 0) $ such that
	\begin{equation}\label{eq:channelAform}
	(S_2, h_2, 0)  (K, l, \alpha) (S_1, h_1, 0) = (A , 0, \Lambda),
	\end{equation}
	where $ A $ is a real skew-symmetric matrix ($ A^T = -A $) and $ \Lambda $ is  a real diagonal matrix;
	
	(3) if the matrix $ -K^T J K J $ has no positive real eigenvalues, there exist two in\-homo\-geneous  symplectic transformations $ (S_1, h_1, 0) $ and $ (S_2, h_2, 0) $ such that
	\begin{equation}\label{eq:channelDAform}
	(S_2, h_2, 0)  (K, l, \alpha) (S_1, h_1, 0) = (D A , 0, \Lambda),
	\end{equation}
	where $ A $ is a real skew-symmetric matrix ($ A^T = -A $), $ D $ is defined by formula \eqref{eq:matD} and $ \Lambda $ is  a real diagonal matrix.
\end{theorem}

\begin{demo}
	First of all, let us note that by applying elements of the form $ (I_{2n}, h, 0) $ of the translation subgroup of the in\-homo\-geneous symplectic group one obtains $ (I_{2n}, h_2, 0)  (K, l, \alpha) (I_{2n}, h_1, 0) = (K ,  h_1 + l + h_2, \alpha) $.
	Thus, one could transform an arbitrary channel $ (K, l, \alpha) $ into channel $ (K, 0, \alpha) $, closing arbitrary $ h_1 $ and $ h_2 $ satisfying  $ h_1 + l + h_2 = 0 $. So now one should simplify a channel of the form $ (K, 0, \alpha) $ by symplectic transforms: $ (S_2, 0, 0) (K, 0, \alpha) (S_1, 0, 0) = (S_1 K S_2, 0,  S_2^T \alpha S_2) $.

	 By multiplying \eqref{eq:GaussCond} on both sides by a real vector $ v $ we obtain $ v^T \alpha v \geqslant 0 $. As $ \alpha $ is real and symmetric it means that it is non-negative definite. Thus, by the Williamson theorem one could take $  S_2 $ such that $ S_2^T \alpha S_2 = \Lambda $, where $ \Lambda $ is a diagonal matrix \cite[Lemma 12.12]{Holevo10},  \cite[App.~6]{Arnold13}, \cite{Williamson39}. So one could simplify the form of the matrix $ X = K S_2 $ by choosing the matrix $ S_1 $, which could be done with our results from the previous section.
	 
	 Let us note that $  -X^T J X J =  -S_2^T K^T J K S_2 J = S_2^T (-K^T J K J) (S_2^T)^{-1}$. Hence the eigenvalues of the matrix $ -X^T J X J $ are the same as the ones of $ -K^T J K J $. Then form \eqref{eq:channelDRform} is a direct consequence of \eqref{eq:decompSDR}, one could just take $ S_1 = S^{-1} $, where $ S $ is from \eqref{eq:decompSDR}. Similarly, \eqref{eq:channelAform} and \eqref{eq:channelDAform} are direct consequences of \eqref{eq:decompSA} and \eqref{eq:decompSDA}, respectively.  \qed
\end{demo}

In the one-mode ($ n=1 $) case one could obtain by direct computation that $ -K^T J K J = \det(K) I_{2} $.
For the non-degenerate case either $  \det(K)  > 0 $ or $  \det(K)  < 0 $. Hence, the matrix $ -K^T J K J $ has either both positive or both negative real eigenvalues. Thus, the conditions of either paragraph (2) 
or paragraph (3) of theorem \ref{th:applToBGQC} are satisfied. They correspond to the cases B)-C) and D) in Holevo's classification \cite{Holevo07}, respectively.

An arbitrary real $ 2n \times 2n $-matrix is defined by $ 4 n^2 $ real parameters, a symplectic matrix is defined by $ n (2n+1) $ real parameters. It suggests that the matrix $ K $ could be transformed in the form parameterized by $ 4 n^2 - n (2n+1) = n (2n-1) $ real numbers. A skew-symmetric real matrix has $  n (2n-1)  $ real parameters, while the symmetric one has $  n (2n+1)  $  real parameters. In this sense general form \eqref{eq:channelDRform} is not optimal and should be further simplified to the forms like  \eqref{eq:channelAform} and \eqref{eq:channelDAform} if possible. 

Moreover, as paragraphs (2) and (3) are based on theorem \ref{th:main} the matrix $ A $ could be calculated in terms of the primary square root, which is obtained by most of computational methods \cite[Ch. 6]{Higham08}. And paragraph (1) is based on theorem \ref{th:main0}, where the Hamiltonian square root of the skew-Hamiltonian matrix occurs, but at the beginning of the proof of theorem \ref{th:main} we have shown that any primary square root of the skew-Hamiltonian matrix must be skew-Hamiltonian. This is another reason for usage of paragraphs (2) and (3) rather than wider applicable paragraph (1). This could be the reason to prefer forms \eqref{eq:channelAform} and \eqref{eq:channelDAform} to the Wolf canonical forms in numerical calculations, because the latter ones are based on the numerically-unstable Jordan decomposition \cite[Sec.~7.8]{Franklin12} as opposite to primary square root computation for which stable methods exist in both complex \cite[Ch. 6]{Higham08}, \cite{Cross74} \cite{Higham97} and real \cite{Higham87} arithmetic.

\section{Conclusions}

We have obtained several analogs of polar decomposition in theorems \ref{th:main0}, \ref{th:main}  and corollaries \ref{cor:decomRDS}--\ref{cor:last}. Then we have applied these results to transformation of the bosonic Gaussian channels by inhomogeneous  symplectic transforms.

The Holevo classification separates the cases B) and C) reflecting the fact that for a given matrix $ K $ the presence of zero diagonal values of $ \Lambda $ could be prohibited by condition \eqref{eq:GaussCond}. So it is interesting for further study to generalize this finer classification on the $ n $-mode case. 

Only for a specific sign of the real eigenvalues of the matrix $ -K^T J K J $ the Gaussain channels are reduced by theorem \ref{th:applToBGQC} to ''optimal and compatible'' (in the sense discussed at the end of Sec.~\ref{sec:bosChan}) forms \eqref{eq:channelAform}--\eqref{eq:channelDAform}. So it is natural to ask if it is possible to obtain similar forms for arbitrary signs of the real eigenvalues.

Both in \cite{Wolf2008} and in our article the case of the degenerate  matrix  $ K $ was out of the range of the study. Thus, it is also a natural candidate for future development.

\section*{Acknowledgments}
The author thanks A.\,S.~Holevo and A.\,N.~Pechen for fruitful discussion leading to posing the problems considered in our work.

This work is supported by the Russian Science Foundation under grant 19-11-00320.

\end{document}